\documentstyle[twocolumn,aps,psfig]{revtex}
\tighten
\def\bk{{\bf k}}

\def\br{{\bf r}}
\def\brp{{\bf r}^\prime}
\def\ha{{\hat{\alpha}}}
\def\hb{{\hat{\beta}}}

\begin{document}
\draft
\title{Projected Wavefunctions and High Temperature Superconductivity}
\author{ Arun Paramekanti, Mohit Randeria and Nandini Trivedi }
\address{Department of Theoretical Physics, Tata Institute of Fundamental
Research, Mumbai 400005, India \\ } 
\address{ 
\begin{minipage}[t]{6.0in}
\begin{abstract} 
We study the Hubbard model 
with parameters relevant to the cuprates, using variational 
Monte Carlo with projected $d$-wave states.
For doping $0 < x \protect\lesssim 0.35$ we obtain a
superconductor whose order parameter tracks the
observed nonmonotonic $T_c(x)$. 
The variational parameter $\Delta_{\rm var}(x)$ scales with
the $(\pi,0)$ ``hump'' and $T^\ast$ seen in photoemission. 
Projection leads to incoherence in the spectral function 
and from the {\it singular} behavior of its moments we obtain the 
nodal quasiparticle weight $Z \sim x$ though the Fermi velocity remains
finite as $x\to 0$.
The Drude weight $D_{\rm low}$ and superfluid
density are consistent with experiment and $D_{\rm low}\!\sim\! Z$. 
\end{abstract} 
\pacs{PACS numbers:
74.20.-z, 74.20.Fg, 74.25.-q, 74.72.-h} 
\end{minipage}} 
\date{\today}
\maketitle

\narrowtext
%==============================================================================

Strong correlations are essential to understand $d$-wave
high temperature superconductivity in doped Mott insulators
\cite{anderson}. 
The no-double occupancy constraint arising from strong correlations 
has been treated within two complementary approaches.
Within the gauge theory approach \cite{gauge}, which is valid at all
temperatures, the constraint necessitates the inclusion
of strong gauge fluctuations.
Alternatively, the constraint can be implemented exactly at $T=0$
using the variational Monte Carlo (VMC) method.
Previous variational studies \cite{rice,gros,shiba,becca} have
focussed primarily on the energetics of competing states.

In this letter, we revisit projected wavefunctions of the form
proposed by Anderson in 1987 \cite{anderson}. We compute physically 
interesting correlations using VMC and 
show that projection leads to loss of coherence.
We obtain information about 
low energy excitations from the singular behavior of moments
of the occupied spectral function. 
Remarkably, our results for various observables
are in semi-quantitative agreement with experiments on the cuprates.
We also make qualitative predictions for the doping ($x$) dependence
of correlation functions in projected states 
(for $x \ll 1$) from general arguments which are largely independent of the
detailed form of the wavefunction and the Hamiltonian.

%%%%%%%%%%%%%%%%%%%%%%%%%%%%%%%%%%%%%%%%%%%%%%%%%%%%%%%%%%%%%%%%%%%%%%%%%%%%

We use the Hubbard Hamiltonian
${\cal H} = {\cal K} + {\cal H}_{\rm int}$.
The kinetic energy 
${\cal K} = \sum_{\bk,\sigma} \epsilon(\bk)c_{\bk\sigma}^{\dag}c_{\bk\sigma}$
with $\epsilon(\bk) = -2t\left(\cos k_x + \cos k_y \right) 
+ 4t^\prime \cos k_x \cos k_y$ the dispersion
on a square lattice with nearest ($t$) and next-near ($t^\prime$) hopping.
The on-site repulsion is
${\cal H}_{\rm int} =  U \sum_{\br} n_{\uparrow}(\br) n_{\downarrow}(\br)$
with $n_{\sigma}(\br) = c_{\sigma}^{\dag}(\br) c_{\sigma}(\br) $.
We work in the strongly correlated regime where $J = 4t^2/U,\ t^\prime  
\lesssim t \ll U$ near half filling: $n = 1-x$ with the hole doping $x \ll 1$.
We choose $t^\prime = t/4$, $t = 300$meV and $U= 12 t$, so that 
$J = 100$meV consistent with neutron data on cuprates.

We describe the ground state by the wavefunction
\begin{equation}
\vert \Psi_0 \rangle = \exp(iS){\cal P} \vert \Psi_{\rm BCS} \rangle.
\label{wavefn}
\end{equation}
Here $\vert \Psi_{\rm BCS} \rangle = \left(\sum_\bk \varphi(\bk) 
c_{\bk\uparrow}^{\dag}c_{-\bk\downarrow}^{\dag}\right)^{N/2} \vert 0 \rangle$
is the $N$-electron d-wave BCS wave function with 
$\varphi(\bk) = v_\bk/u_\bk 
= \Delta_\bk/[\xi_\bk + \sqrt{\xi_\bk^2 + \Delta_\bk^2}]$.
The two variational parameters $\mu_{\rm var}$ and $\Delta_{\rm var}$ 
determine $\varphi(\bk)$ through 
$\xi_\bk = \epsilon(\bk) - \mu_{\rm var}$ and
$\Delta_\bk = \Delta_{\rm var}\left(\cos k_x - \cos k_y\right)/2$.
The projector ${\cal P} = \prod_{\br}\left(
1 - n_{\uparrow}(\br) n_{\downarrow}(\br)\right)$ in Eq.~(\ref{wavefn})
eliminates all configurations with double occupancy, as appropriate 
for $U \to \infty$. 

The unitary transformation $\exp(iS)$ includes
double occupancy perturbatively in $t/U$ \cite{macdonald}. 
The transformation $\exp(-iS){\cal H}\exp(iS)$
is well known to lead to the tJ Hamiltonian \cite{macdonald}.
We emphasize that our approach effectively
transforms {\it all} operators, and not only ${\cal H}$.
Using $\langle\Psi_0\vert {\cal Q} \vert\Psi_0\rangle
= \langle\Psi_{\rm BCS}\vert {\cal P}\tilde{{\cal Q}}
 {\cal P} \vert \Psi_{\rm BCS} \rangle$ with
$\tilde{{\cal Q}}= \exp(-iS){\cal Q}\exp(iS)$, it follows that
incorporating $\exp(iS)$ in the wavefunction (1) is equivalent to
transforming ${\cal Q}\! \to\! \tilde{{\cal Q}}$ and working with 
fully projected states. 

%%%%%%%%%%%%%%%%%%%%%%%%%%%%%%%%%%%%%%%%%%%%%%%%%%%%%%%%%%%%%%%%%%%%%%%%%%%%

Using standard VMC techniques \cite{vmc}
we compute equal-time correlators
in the state $\vert \Psi_0 \rangle$.
The two variational parameters are determined by minimizing
$\langle \Psi_0 \vert{\cal H}\vert \Psi_0 \rangle/
\langle \Psi_0 \vert \Psi_0 \rangle$ at each $x$. 
The doping dependence of the resulting
$\Delta_{\rm var}(x)$ is shown in Fig.~1(a). Varying input parameters
in ${\cal H}$ we find that the scale for $\Delta_{\rm var}$ is mainly 
determined by $J = 4t^2/U$.
We show below that $\Delta_{\rm var}$ is {\it not} proportional
to the SC order parameter, in contrast to BCS theory,
and also argue that it is {\it not} equal to the spectral gap. 
On the other hand, we find that the optimal
$\mu_{\rm var}(x) \approx \mu_{\small 0}(x)$, the
chemical potential of {\it non}interacting electrons described by ${\cal K}$.
However, $\mu_{\rm var}$ is quite different \cite{long}
from the chemical potential
$\mu = \partial \langle {\cal H} \rangle/ \partial N$,
where $\langle \cdots \rangle$ denotes the expectation value
in the optimal, normalized ground state.

{\bf Phase Diagram:} We now show that the wavefunction (\ref{wavefn})
is able to describe three phases: a resonating valence bond (RVB) insulator, 
a $d$-wave SC and
a Fermi liquid metal.  To establish the $T=0$ phase diagram
we first study off-diagonal long range order (ODLRO) using $F_{\alpha,\beta}
(\br - \brp) =
\langle c_{\uparrow}^{\dag}(\br) c_{\downarrow}^{\dag}(\br + \ha) 
c_{\downarrow}(\brp) c_{\uparrow}(\brp + \hb) \rangle$, 
where $\ha,\hb = {\hat x}, {\hat y}$.
We find that $F_{\alpha,\beta} \to \pm \Phi^2$ for large
$\vert \br - \brp \vert$, with $+$ ($-$) sign obtained for 
$\ha \parallel$ ($\perp$) to $\hb$, indicating $d$-wave SC.
As seen from Fig.~1(b) the order parameter
$\Phi(x)$ is {\it not} proportional to $\Delta_{\rm var}(x)$, 
and is nonmonotonic. $\Phi$ vanishes linearly in $x$ as $x \to 0$
as first noted in ref.~\cite{gros}, even though $\Delta_{\rm var} \ne 0$. 
We argue \cite{footnote1} that $\Phi \sim x$ 
is a general property of projected SC wavefunctions.
The local fixed number constraint imposed by ${\cal P}$
at $x = 0$ leads to large quantum phase fluctuations that destroy SC order.
The non-SC state at $x = 0$ is an insulator with a vanishing Drude weight, 
as shown below.
The system is a SC in the doping range $0< x < x_c \simeq 0.35$
with $\Phi \ne 0$. For $x > x_c$, $\Phi = 0$
and the wavefunction $\Psi_0$ for $\Delta_{\rm var} = 0$ 
describes a normal Fermi liquid.
In the remainder of this paper we will study $0 \le x \le x_c$. 
%%%%%%%%%%%%%%%%%%%%%%%%%%%%%%%%%%%%%%%%%%%%%%%%%%%%%%%%%%%%%%%%%%%%%%%%%%%%%%%
\begin{figure}
\begin{center}
\vskip-2mm
\hspace*{0mm}
\psfig{file=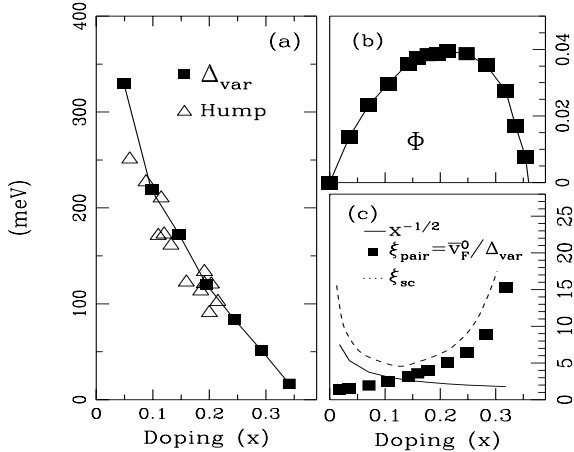,height=2.5in,width=3.1in,angle=0}
\vskip0mm
\caption{(a): The variational parameter $\Delta_{\rm var}$ (filled
squares) and the $(\pi,0)$ hump scale (open triangles)
in ARPES \protect\cite{campuzano} versus doping. (b): 
Doping dependence of the $d$-wave SC order parameter $\Phi$.
Solid lines in (a) and (b) are guides to the eye.
(c): The coherence length
$\xi_{\rm sc} \protect\geq {\rm max} (\xi_{\rm pair},1/\sqrt{x})$.}
\label{fig1}
\end{center}
\end{figure}
%%%%%%%%%%%%%%%%%%%%%%%%%%%%%%%%%%%%%%%%%%%%%%%%%%%%%%%%%%%%%%%%%%%%%%%%%%%%%%%

{\bf Coherence Lengths}:
We must carefully distinguish between various `coherence lengths', which 
are the same in BCS theory, but are very different in strongly correlated 
SC's. The internal pair wavefunction $\varphi(\bk)$ defines a pair-size
$\xi_{\rm pair} \sim \bar{v}^0_F/\Delta_{\rm var}$,
where $\bar{v}^0_F$ is the bare average Fermi velocity. 
Projection does not affect the pair-size much, and $\xi_{\rm pair}$
remains finite at $x = 0$, where
it defines the range of singlet bonds in the
RVB insulator \cite{anderson}. 

A second important length scale is the inter-hole spacing $1/\sqrt{x}$.  
At shorter distances there are no holes and the system effectively looks 
like the $x=0$ insulator. Thus the superconducting coherence length
$\xi_{\rm sc}$ must necessarily satisfy 
$\xi_{\rm sc} \ge \max\left(\xi_{\rm pair},1/\sqrt{x}\right)$.
This bound implies that $\xi_{\rm sc}$ 
must diverge both in the insulating limit $x \to 0$ 
(see also Refs.~\cite{gauge}) and the metallic limit $x \to x_c^{-}$,
but is small at optimality; see Fig.~1(c).
This non-monotonic behavior of $\xi_{\rm sc}(x)$ should be 
checked experimentally.

%%%%%%%%%%%%%%%%%%%%%%%%%%%%%%%%%%%%%%%%%%%%%%%%%%%%%%%%%%%%%%%%%%%%%%%%%%%%%%

{\bf Momentum Distribution}: From gray-scale plots like Fig.~2 we see 
considerable structure in the momentum distribution $n(\bk)$ over the
entire range $0 \le x \le x_c$. One cannot define a Fermi
surface (FS) since the system is a SC (or an insulator).
Nevertheless, for all $x$, the $\bk$-space loci (i) on which $n(\bk) = 1/2$
and (ii) on which $\vert \nabla_\bk n(\bk) \vert$ is maximum, are both
quite similar to the corresponding {\it non}interacting FS's. 
For $t^\prime = t/4$, 
the FS is hole-like for $x\protect\lesssim 0.22$, and electron-like 
for overdoping \cite{fujimori}.

We next exploit the fact that moments of dynamical correlations
can be expressed as equal-time correlators, calculable
in our formalism. Specifically, we look at
$M_\ell(\bk) = \int_{-\infty}^\infty d\omega \omega^\ell f(\omega) 
A(\bk,\omega)$, where $A(\bk,\omega)$ is the one-particle spectral function
and, at $T=0$, $f(\omega) = \Theta(-\omega)$. 
We calculate $M_0(\bk) = n(\bk)$ and the first moment \cite{footnote2}
$M_1(\bk) = \langle c_{\bk\sigma}^{\dag} [{\cal H}, c_{\bk\sigma}]\rangle$
to obtain important information about the spectral function.

%%%%%%%%%%%%%%%%%%%%%%%%%%%%%%%%%%%%%%%%%%%%%%%%%%%%%%%%%%%%%%%%%%%%%%%%%%%%%%%

{\bf Nodal Quasiparticles}: Fig.~2(c) shows that 
along $(0,0)$ to $(\pi,\pi)$ $n(\bk)$
has a jump discontinuity. This implies the existence of
gapless quasiparticles (QPs) observed by 
angle resolved photoemission spectroscopy                        
(ARPES) \cite{kaminski}. 
The spectral function along the diagonal thus has the low energy form:
$A(\bk,\omega) = Z \delta(\omega - \tilde{\xi}_k) + A_{inc}$,
where $\tilde{\xi}_k = v_F(k - k_F)$ is the QP dispersion and
$A_{inc}$ the smooth incoherent part. 
We estimate the nodal $k_F(x)$ from the location of the discontinuity
and the QP weight $Z$ from its magnitude.
While $k_F(x)$ has weak doping dependence,
$Z(x)$ is shown in Fig.~2(d), with
$Z\sim x$ as the insulator is approached \cite{footnote3}. 

Projection leads to a suppression of $Z$ from
unity with the incoherent weight $(1-Z)$
spread out to high energies.
We infer large incoherent linewidths as follows.
(a) At the ``band bottom'' $n(\bk = (0,0)) \simeq 0.85$ (for $x = 0.18$) 
implying that 15\% of the spectral weight must have
spilled over to $\omega > 0$. (b) Even {\it at} $k_F$, the first moment 
$M_1$ lies significantly below $\omega = 0$ (defined by the chemical
potential $\mu = \partial \langle {\cal H} \rangle/ \partial N$);
see Fig.~3(a).  

The moments are dominated by the high energy incoherent 
part of $A(\bk,\omega)$, but their {\em singular behavior is determined
by the gapless coherent} QPs. Specifically, along the zone
diagonal $M_1(\bk) = Z \tilde{\xi}_k \Theta(-\tilde{\xi}_k) + {\rm
smooth\ part}$.  Thus its slope $dM_1(\bk)/dk$ has a
discontinuity of $Z v_F$ at $k_F$, as seen in Fig.~3(a),
and may be used to estimate \cite{footnote4} 
the nodal Fermi velocity $v_F$.
As seen from Fig.~3(b), $v_F(x)$ is reduced from
its bare value $v^0_F$ and is weakly doping
dependent, consistent with the ARPES estimate\cite{kaminski} of
$v_F \approx 1.5 eV$-$\AA$ in Bi$_2$Sr$_2$CaCu$_2$O$_{8+\delta}$ (BSCCO).

As $x \to 0$, $Z = [1 - \partial \Sigma^\prime/ \partial\omega]^{-1} \sim x$, 
while
$v_F(x)/v^0_F = Z [1 + (v^0_F)^{-1} \partial \Sigma^\prime/ \partial k]$ 
is weakly $x$-dependent. This implies a $1/x$ divergence in the
$\bk$-dependence of the self energy $\Sigma$
on the zone diagonal, which could be tested by ARPES.

Within slave boson mean field theory (MFT) \cite{anderson}, we find \cite{long} 
$Z^{\rm sb}\!\!=\!\!x$ in Fig.2(d) and 
$v^{\rm sb}_F(x)$ shown in Fig.3(b), both
systematically smaller than the corresponding VMC results. Thus
the holons and spinons must be 
partially bound by gauge fluctuations beyond MFT.

%%%%%%%%%%%%%%%%%%%%%%%%%%%%%%%%%%%%%%%%%%%%%%%%%%%%%%%%%%%%%%%%%%%%%%%%%%%%%%%

{\bf Moments Along $(\pi,0) \to (\pi,\pi)$}: 
The moments $n(\bk)$ and $M_1(\bk)$ near
$\bk = (\pi,0)$ are not sufficient to estimate the
SC gap, however they give insight into the nature of the
spectral function. As seen from Fig.~4(a) $n(\bk)$ is
much broader than that for the unprojected $\vert \Psi_{\rm BCS}\rangle$. 
For $\bk$'s near $(\pi,\pi)$,
which correspond to high energy, unoccupied ($\omega >0$) states in
$\vert \Psi_{\rm BCS}\rangle$ we see a significant build up of spectral
weight transferred from $\omega < 0$. Correspondingly, we see
loss of spectral weight near $(\pi,0)$. We thus infer
large linewidths at all $\bk$'s, also seen 
from the large $\vert M_1(\bk)\vert$ (in Fig.~4(b))
relative to the BCS result.

The transfer of weight over large energies inferred 
above suggests that projection pushes spectral weight to
$\vert \omega \vert < \Delta_{\rm var}$,
the gap in unprojected BCS theory. We thus expect that 
the true spectral gap $E_{\rm gap} < \Delta_{\rm var}$ and
that the coherent QP near $(\pi,0)$
must then reside at $E_{\rm gap}$ in order
to be stable against decay into incoherent excitations. 
It is then plausible that $\Delta_{\rm var}$, the large gap before projection, 
is related to an incoherent feature in $A(\bk,\omega)$ near 
$(\pi,0)$ after projection \cite{footnote5}. Indeed, comparing 
$\Delta_{\rm var}$ with the $(\pi,0)$ ``hump''
feature seen in ARPES\cite{campuzano}, we find
good agreement in the magnitude as well as doping
dependence; see Fig.~1(a).

%%%%%%%%%%%%%%%%%%%%%%%%%%%%%%%%%%%%%%%%%%%%%%%%%%%%%%%%%%%%%%%%%%%%%%%%%%%%%%%

{\bf Optical spectral weight}: The optical conductivity sum rule
states that $\int_{0}^\infty d\omega Re \sigma(\omega)$ $=$ $\pi 
\sum_{\bk} m^{-1}(\bk) n(\bk)$ $\equiv \pi D_{\rm
tot}/2$ where $m^{-1}(\bk) = $ $\left(\partial^2
\epsilon(\bk)/ \partial\bk_x \partial\bk_x \right)$ 
is the {\it non}interacting mass tensor (we set $\hbar=c=e=1$).
The {\it total} optical spectral weight $D_{\rm tot}(x)$ 
is plotted in Fig.~5(a) and seen to be non-zero even at $x=0$, 
since the infinite cutoff in the integral above includes contributions 
from the ``upper Hubbard band''.

A physically more interesting quantity is the {\it low frequency}
optical weight, or Drude weight, $D_{\rm low}$ where the
upper cutoff extends above the scale of $t$ and $J$, but is much
smaller than $U$. This is conveniently defined \cite{long} by the
response to an external vector potential: $D_{\rm low} = \partial^2
\langle {\cal H}_A \rangle_A /\partial A^2$.  Here the subscript
on ${\cal H}$ denotes that $A$ enters the kinetic energy via a Peierls
minimal coupling, and that on the expectation value denotes the corresponding
modification of $\exp(iS)$ in Eq.~(\ref{wavefn}). The
Drude weight $D_{\rm low}(x)$, plotted in Fig.~5(a), 
vanishes linearly as $x \to 0$, which can be argued to be a general
property of projected states \cite{long}. 
This also proves, following Ref.~\cite{millis}, that 
$\vert \Psi_0 \rangle$ describes an insulator at $x=0$.
Both the magnitude and doping dependence of $D_{\rm low}(x)$ are 
consistent with optical data on the cuprates \cite{cooper}.  

Motivated by our results on the nodal $Z(x)$ and $D_{\rm low}(x)$,
we make a parametric plot of these two quantities in Fig.~5(b) and 
find that $D_{\rm low} \sim Z$ over the entire doping range,
a prediction which can be checked by comparing optics and ARPES \cite{shen}
on the cuprates.

%%%%%%%%%%%%%%%%%%%%%%%%%%%%%%%%%%%%%%%%%%%%%%%%%%%%%%%%%%%%%%%%%%%%%%%%%%%%%%%

{\bf Superfluid Density:} 
The Kubo formula for the superfluid stiffness $D_s$ can be written as 
$D_s = D_{\rm low} - \Lambda^{T}$, where $D_{\rm low}$ is the
diamagnetic response, and $\Lambda^{T}$ is the transverse current-current
correlator. Using the spectral representation
for $\Lambda^{T}$ it is easy to see that \cite{bound}
$\Lambda^{T} \geq 0$ implying $D_s \leq D_{\rm low}$. 
Two conclusions follow. First, $D_s \to 0$
as $x\to 0$, consistent with experiments \cite{uemura}. Second,
we obtain a lower bound on the penetration depth
$\lambda_{\rm L}^{}$ defined by 
$\lambda^{-2}_{\rm L} \equiv 4 \pi e^2 D_s/\hbar^2 c^2 d_c$.
Using the calculated $D_{\rm low} \approx 90 meV$ at optimality and 
a mean interlayer spacing $d_c = 7.5 \AA$ (appropriate to BSCCO), we find 
$\lambda_{\rm L}^{}\gtrsim 1350 \AA$, consistent with
experiment \cite{lambda}.

{\bf Conclusions:}
We have shown, within our variational scheme, the $T=0$ evolution of
the system from an undoped insulator to a d-wave 
SC to a Fermi liquid as a function of $x$.
The SC order parameter $\Phi(x)$ is non-monotonic with a maximum at optimal 
doping $x\simeq 0.2$,

%%%%%%%%%%%%%%%%%%%%%%%%%%%%%%%%%%%%%%%%%%%%%%%%%%%%%%%%%%%%%%%%%%%%%%%%%%%%%%%
\begin{figure}
\begin{center}
\vskip-2mm
\hspace*{0mm}
\psfig{file=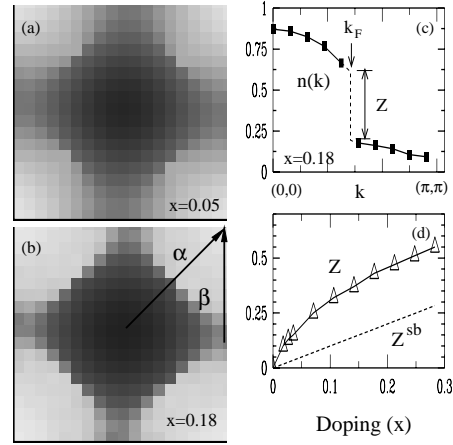,height=2.5in,angle=0}
\vskip3mm
\caption{(a) and (b): Gray-scale plots of $n(\bk)$ (black $\equiv 1$,
white $\equiv 0$) centered at
$\bk=(0,0)$ for $x=0.05$ and $x=0.18$ respectively on a $19\times19+1$ 
lattice, showing very little doping
dependence of the large ``Fermi surface''. (c): 
$n(\bk)$ plotted along the diagonal
direction (indicated as $\alpha$ in Panel (b)), showing the jump
at $k_F$ which implies a gapless nodal quasiparticle of weight $Z$.
(d): Nodal quasiparticle weight $Z(x)$,
compared with the slave boson mean field $Z^{\rm sb}(x)$ (dashed line).}

\label{fig2}
\end{center}
\end{figure}
%%%%%%%%%%%%%%%%%%%%%%%%%%%%%%%%%%%%%%%%%%%%%%%%%%%%%%%%%%%%%%%%%%%%%%%%%%%%%%%
\begin{figure}
\begin{center}
\vskip-4mm
\hspace*{0mm}
\psfig{file=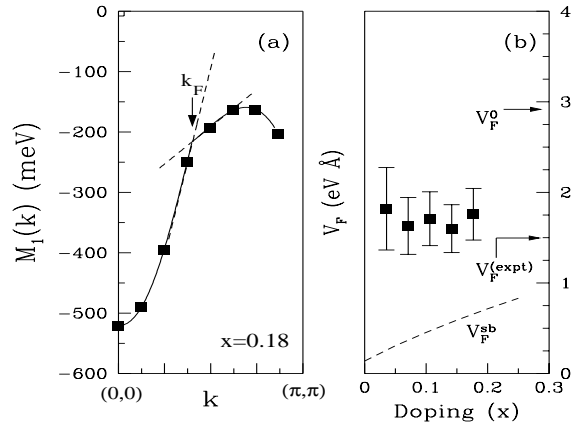,height=2.5in,width=3.1in,angle=0}
\vskip-2mm
\caption{(a): The moment $M_1(\bk)$
along the zone diagonal, with smooth fits for $k < k_F$
and $k > k_F$, showing a discontinuity of $Z v_F$ in its slope
at $k_F$.
(b): Doping dependence of the nodal quasiparticle
velocity obtained from $M_1(\bk)$. Error bars come from
fits to $M_1(\bk)$ and errors in $Z$. Also shown are 
the bare nodal velocity $v^0_F$, the slave boson mean field $v^{\rm sb}_F(x)$ 
(dashed line), and the ARPES estimate $v^{\rm (expt)}_F$
\protect\cite{kaminski}.}
\label{fig3}
\end{center}
\end{figure}
%%%%%%%%%%%%%%%%%%%%%%%%%%%%%%%%%%%%%%%%%%%%%%%%%%%%%%%%%%%%%%%%%%%%%%%%%%%%%%%

\noindent 
suggestive of the experimental trend in $T_c(x)$ from under- to over-doping.
As $x \to 0$ $D_s$ vanishes, while the
spectral gap, expected to scale with $\Delta_{\rm var}$, remains
finite. Thus the underdoped state, with strong pairing and
weak phase coherence, should lead to pseudogap behavior
in the temperature range between $T_c$ (which scales like $\Phi$)
and $T^*$ (which scales like $\Delta_{\rm var}$).

%%%%%%%%%%%%%%%%%%%%%%%%%%%%%%%%%%%%%%%%%%%%%%%%%%%%%%%%%%%%%%%%%%%%%%%%%%%%%%%
\begin{figure}
\begin{center}
\vskip-2mm
\hspace*{0mm}
\psfig{file=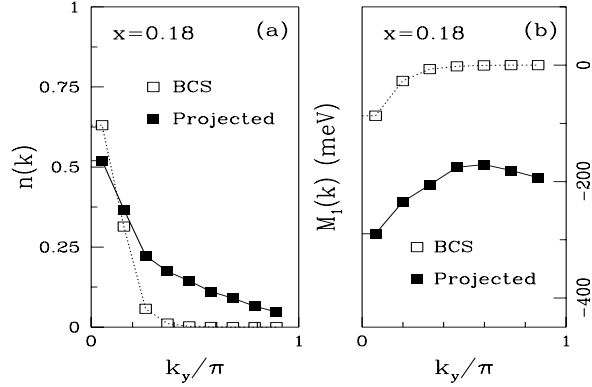,height=2.5in,width=3.1in,angle=0}
\vskip-2mm
\caption{(a): $n(\bk)$ plotted along the $(\pi,0)$-$(\pi,\pi)$
direction (indicated as $\beta$ in Fig.~2(b)) and compared with the 
BCS result. (b): The moment $M_1(\bk)$
plotted along the $(\pi,0)$-$(\pi,\pi)$ direction compared with the BCS 
values.} 
\label{fig4}
\end{center}
\end{figure}

\begin{figure}
\begin{center}
\vskip-4mm
\hspace*{0mm}
\psfig{file=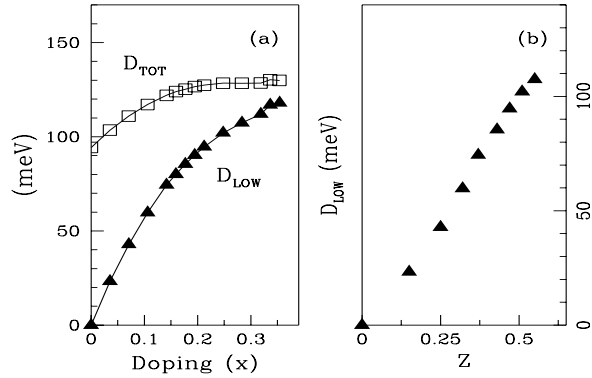,height=2.5in,width=3.1in,angle=0}
\vskip-2mm
\caption{(a): Doping dependence of the total ($D_{\rm tot}$) and 
low energy ($D_{\rm low}$) optical spectral weights
(b): The optical spectral weight $D_{\rm low}$ versus 
the nodal quasiparticle weight $Z$.} 
\label{fig5}
\end{center}
\end{figure}

%%%%%%%%%%%%%%%%%%%%%%%%%%%%%%%%%%%%%%%%%%%%%%%%%%%%%%%%%%%%%%%%%%%%%%%%%%%%%
{\bf Acknowledgements}: 
We thank H.R. Krishnamurthy, P.A. Lee, M.R. Norman and J. Orenstein 
for helpful comments on an earlier draft.
M.R. was supported in part by the DST through the Swarnajayanti scheme.
%==============================================================================
%                      REFERENCES
%==============================================================================


\begin{references}
\bibitem{vmc}
Since the pair wavefunction $\varphi(\bk)$ is singular on the zone
diagonal we use ``tilted'' lattices \cite{rice} 
with periodic boundary conditions.
Most results were obtained on a $15 \times 15 + 1$ site lattice,
with some on a $19 \times 19 + 1$ lattice.
Typically we used 5000 equilibriation sweeps, and averaged over
500-1000 configurations from 5000 sweeps.
In all the VMC results presented here, the error bars are of the 
size of the symbol, unless explicitly shown.
\bibitem{footnote1}
To put a pair of electrons on a link near $\br$,
after removing them from a link near the origin, requires two vacancies
near $\br$. Thus $F_{\alpha,\alpha}(\br) \sim x^2$ and $\Phi \sim x$
as $x \to 0$.
\bibitem{footnote2}
The Fourier transform of $M_1(\bk)$ is $M_1(\br-\brp) = - t_{\br\brp}
\langle c^\dag_\sigma (\br) c_\sigma (\brp)\rangle+ U \langle 
c^\dag_\sigma(\br)  
c_\sigma(\brp) n_{\overline{\sigma}}(\brp) \rangle$.  Since the second
term is ${\cal O}(U)$ we need $S$ in Eq.~(\ref{wavefn}) to ${\cal O}(t/U)^2$ 
to obtain $M_1(\bk)$ correctly to ${\cal O}(J)$.
\bibitem{footnote3}
Relating $Z$ to the amplitude of the asymptotic power-law decay in
$\langle c_{\sigma}^{\dag}(\br) c_{\sigma}(\brp) \rangle$ we can show that
$Z \sim x$ for projected states \cite{long}.
\bibitem{footnote4}
For unprojected BCS wavefunctions we can estimate
$v_F$ with $5\%$ accuracy using this method on finite systems.
\end{references}
\end{document}